\documentstyle[12pt]{article}
\newcounter{eqnn} \newcounter{eqs} \newcounter{secn}
\newcounter{subn}
\def\sec{\addtocounter{secn}{1}\setcounter{subn}{0}
\setcounter{eqs}{0}$\bf \thesecn $}

\def\eq{\addtocounter{eqnn}{1}\;\;\;(\theeqnn)} 
\def\lae{\;^{<}_{\sim} \;} \def\gae{\; ^{>}_{\sim} \;}

\def\s2t{Sin^{2}\theta_{w}}

\begin{document}

\begin{titlepage}

\pagestyle{empty} \begin{center} {\Large \bf  
}\end{center}              \begin{flushright}  HELSINKI\\              
March 1998 \end{flushright}              \vfill              
\begin{center} {\LARGE   
Spontaneous 
Discrete Symmetry Breaking During Inflation
 and the NMSSM Domain Wall Problem      
\\ } \end{center}
\vfill \begin{center} {\bf John McDonald
\begin{footnote}{e-mail:
mcdonald@phcu.helsinki.fi}\end{footnote} }\\       \vspace
{0.1in}     Dept. of Physics,\\P.O.Box 9,\\University of
Helsinki,\\ FIN-00014 Helsinki,\\FINLAND \\ \vfill
\end{center} \newpage \begin{center} {\bf Abstract}
\end{center}  

              The Next to Minimal Supersymmetric Standard
Model (NMSSM), proposed as a solution of the 
$\mu$ problem of the Minimal Supersymmetric Standard Model, 
 has a discrete $Z_{3}$ symmetry 
which is spontaneously broken at the electroweak phase transition, 
resulting in a cosmological domain wall problem. 
In most cases this domain wall problem cannot 
be solved by explicit $Z_{3}$ breaking without 
introducing supergravity tadpole 
corrections which destabilize the weak scale hierarchy.
Here we consider the possibility of solving the domain 
wall problem of the NMSSM 
via spontaneous discrete symmetry breaking occuring during inflation. 
For the case where the discrete symmetry breaking 
field has renormalizible couplings 
to the NMSSM fields, we find that the 
couplings must be less than $10^{-5}$
if the reheating temperature is larger than $10^{7}GeV$, 
but can be up to $10^{-3}$
 for reheating temperatures of the order
of the electroweak phase transition temperature. 
For the case of non-renormalizible couplings, we present 
a model which can solve the domain wall 
problem for large reheating temperatures without 
requiring any very small coupling constants.
 In this model the
domain walls are eliminated by a pressure coming from their interaction 
with a coherently oscillating scalar field whose phase is 
fixed during inflation.
 This oscillating scalar field 
typically decays after the electroweak phase 
transition but before nucleosynthesis, leaving no additional
 $Z_{3}$ symmetry breaking in the zero-temperature theory. 

\end{titlepage}

{\bf \sec. Introduction.} 

              The Minimal Supersymmetric extension of the Standard Model 
(MSSM) has become widely regarded as the most probable theory of 
physics beyond the Standard Model (SM), due to its ability to 
protect the weak scale from quadratically divergent radiative corections 
\cite{nilles}. 
However, the MSSM requires, in order to give an expectation
 value to both Higgs doublets, the introduction of a SUSY mass parameter 
for the Higgs doublets, $\mu$, which has no obvious reason to be of the 
order of $m_{W}$, although some mechanisms have been suggested in the 
context of supergravity 
models \cite{gm}. (We will refer to this as the $\mu$ problem). 
One particularly simple and attractive solution of the $\mu$ problem 
is to extend the MSSM to the $Z_{3}$ symmetric Next to Minimal
Supersymmetric Standard Model (NMSSM) \cite{nilles,nmssm,ets}.
 This adds a scalar superfield $N$ to the MSSM whose
expectation value provides an effective $\mu$ term for the
Higgs doublets. The $\mu$ term is naturally of the order of
the weak scale so long as the $Z_{3}$ symmetry is imposed.
However, the $Z_{3}$ symmetry introduces a cosmological problem
in the case where the reheating temperature after
inflation is large compared with the weak scale, 
namely the formation of stable
domain walls \cite{dwp,dwp2} due to sponanteous 
breaking of the $Z_{3}$ symmetry at the 
electroweak phase transition \cite{saw}.

         In order to eliminate these domain walls, a 
$Z_{3}$ symmetry breaking which has the same phase over 
the region corresponding to the 
observable Universe must be introduced into the effective weak scale theory.
However, any $explicit$ $Z_{3}$ symmetry breaking ($even$ $if$ due to 
Planck-scale suppressed non-renormalizible operators) 
will almost certainly destabilize the weak scale 
by allowing quadratically divergent 
gauge singlet tadpole diagrams which appear in the full supergravity theory 
\cite{saw,bag}. (However, for the special case where the $Z_{3}$ is
embedded in a gauged R-symmetry, such corrections can be avoided \cite{abel}).

                   The obvious alternative is to consider $spontaneous$ 
$Z_{3}$ symmetry breaking due to a new gauge singlet scalar field, $S$, 
whose initial value is fixed during inflation. 
 This ensures that the resulting discrete symmetry breaking phase 
is the same over the whole observable Universe. 
In this paper we will discuss whether such a possibility can be 
implemented in a natural way, in the sense of not requiring extremely 
small coupling constants. 
We will consider two different approaches. 
One approach is based on coupling $S$
 directly to the NMSSM via renormalizible couplings. The second is 
based on having only non-renormalizible
couplings of $S$ to the NMSSM sector, suppressed by powers of 
the Planck scale or some intermediate mass scale.

        We first outline the cosmological scenario we have in mind.
We will consider the simplest scenario, in 
which there is an initial period of inflation
due to the energy density of an inflaton field, which 
subsequently oscillates coherently about the 
minimum of its potential until it decays, leaving the Universe 
radiation dominated at a reheating
temperature $T_{R}$ \cite{hm2}. $T_{R}$ should be low 
enough not to thermally regenerate
gravitinos \cite{gravb,eu} ($T_{R} \lae 10^{8}-10^{9}GeV$ for 
gravitino masses in the range 
$100 GeV$ to $500 GeV$ \cite{gravb2}), 
implying that the value of the
 expansion rate of the Universe at $T_{R}$, 
$H(T_{R})$, should be less than $1GeV$. 
 Between the end of inflation and the beginning of radiation domination
 the Universe is dominated 
by the energy density of the coherent inflaton oscillations. 
An important point to note is that the 
Universe does not "reheat" as such. During the inflaton
 oscillation dominated period the decay 
of the inflaton results in a themal background 
of particles with a temperature 
$T_{r} \approx k_{r} (M_{Pl}HT_{R}^{2})^{1/4}$ (where 
$k_{r} \equiv \left(\frac{9}{5 \pi^{3}} 
\frac{g(T_{R})}{g(T)^{2}}\right)^{1/8}$ is approximately 
equal to 0.4 for temperatures larger than 100 $GeV$),
 the energy density of which eventually comes to dominate the energy density
of the Universe at $T_{R}$ \cite{hm2}. 
The magnitude of the density 
perturbations observed by the Cosmic Background
 Explorer implies that $H \approx 10^{14} GeV$ during inflation, 
assuming that the 
density perturbations are due to quantum fluctuations \cite{cobe}.
Therefore it is possible that $T_{r}$ could exceed $10^{9}GeV$ at the end of 
inflation, even if $T_{R}$ is less than $10^{9}GeV$. 
 In general $T_{r}$ is given by
$${ T_{r} \approx 6.3 
\times 10^{9} k_{r} \left(\frac{H}{100 GeV}\right)^{1/4}
\left(\frac{T_{R}}{10^{9}GeV}\right)^{1/2} GeV      \eq}.$$
Although it might seem that $T_{r} \gae 10^{9}GeV$ would require a 
tighter bound on $T_{R}$ in order to evade the thermal regeneration 
of gravitinos, 
it is straightforward to show that the rapid increase of $H$ 
as a function of $T$ during the 
inflaton oscillation dominated period, $ H \propto T^{4}$, 
ensures that the gravitinos generated thermally 
during this period never exceed those generated around $T_{R}$. 
Thus it is sufficient to impose 
$T_{R} \lae 10^{9}GeV$. On the other hand, we will find 
that the higher temperatures which 
exist during the inflaton oscillation dominated period must be 
taken into account when considering constraints on the couplings 
following from the requirement that the $Z_{3}$ 
symmetry is not thermally restored before the NMSSM domain walls 
form. 

                   It has become clear that in many 
inflation models based on supergravity, the
 scalar fields will obtain order $H^{2}$ SUSY 
breaking mass squared terms as a result of the 
non-zero energy density which exists in the early Universe \cite{hm2,hm}.
 (A possible exception is the case of D-term inflation, depending on the 
details of the model \cite{dti}). 
If such mass squared terms happen to be negative,
then any scalar field with a flat scalar potential will have a
large initial expectation value which is fixed during inflation. 
This is in contrast with the case of potentials having hidden sector 
SUSY breaking mass squared terms, which have a fixed value, $m_{s}^{2}$, 
which is typically of the 
order of $(100GeV)^{2}$. (By "hidden sector" mass terms
 we mean those coming from a supergravity hidden sector breaking 
mechanism \cite{nilles}). 
In this case the fields cannot roll to the 
minimum of their potentials during inflation since $H$ is 
much larger than $m_{s}$ and the fields are overdamped. 
Therefore they will have random initial values.
The role of the $H^{2}$ mass terms in our discussion will be to fix 
the initial values of the fields at the end of inflation 
to be at a non-zero minimum of the scalar potential, with a single 
discrete symmetry breaking 
phase througout the observed region of the Universe. 
It is interesting to
 ask whether the domain wall problem can be solved with random initial
 values of the fields at the end of inflation, 
as in the case without the order $H^{2}$ mass terms, 
but we will not address this question here.

                There are then two possible scenarios 
for the cosmological evolution of the $S$ 
expectation value, depending on the sign of $m_{s}^{2}$.
If this is negative, the 
$S$ expectation value will remain large and 
constant once the hidden sector mass 
squared term comes to dominate the order $H^{2}$ term 
during the expansion of the Universe.
On the other hand, if it is positive, the $S$
 scalar will begin to oscillate coherently 
once $m_{s}^{2}$ dominates the 
$H^{2}$ term. 
If these $Z_{3}$ symmetry breaking 
expectation values can survive down to temperatures less 
than that of the electroweak phase transition, $T_{ew}$, 
(at which the $Z_{3}$ domain walls 
form), then we may be able 
to use the interaction of $<S>$ with the 
NMSSM sector fields to eliminate the 
domain walls. 

                However, we will see that it is not so 
easy to achieve a natural solution of the domain wall problem. 
On the one hand, we have to provide a large enough vacuum 
energy density splitting between the different $Z_{3}$ 
domains to provide sufficient pressure to drive away the domain walls. 
On the other hand, for the case 
of the negative hidden sector mass squared term, 
we have to ensure that the $Z_{3}$ symmetry is not thermally 
restored while at the same time ensuring that the $S$ expectation value
 does not give a large mass to 
the scalars in the NMSSM sector responsible for 
electroweak symmetry breaking. For the case of the 
coherently oscillating $S$ scalar, 
we have to ensure that the $S$ field is not thermalized 
before the electroweak phase transition and that it decays 
after the electroweak phase transition but not so
 late that it disrupts the predictions of standard 
big-bang nucleosynthesis \cite{gravb2}. 

\newpage{\bf \sec. Renormalizible $S$ couplings to the NMSSM sector.}
\newline {\bf \underline{ (a) $m_{s}^{2} < 0$.}}

           We define the transformation
 of a field $\phi$ with charge $Q(\phi)$ under 
a discrete 
symmetry by $ \phi \rightarrow e^{i 2 \pi Q} 
\phi$ ($ -\frac{1}{2} < Q \leq
 \frac{1}{2} $). The simplest extension of the NMSSM 
is then obtained by considering 
the $Z_{3}$ charge of $S$ to be equal to that of $N$, $Q(S) = 1/3$. 
The superpotential of the model is
given by $W = W_{NMSSM} + W_{S}$, where
$${ W_{NMSSM}
= W_{Yukawa} + \lambda_{N} N H_{1} H_{2} - \frac{k}{3} N^{3}     \eq}$$
is the standard $Z_{3}$ symmetric NMSSM superpotential 
\cite{nilles,nmssm,ets,saw} and
$${ W_{S} = \lambda_{SH} S H_{1} H_{2} + \frac{\lambda_{1}}{3} S^{3}
+ \lambda_{2} S^{2}N + \lambda_{3} S N^{2}   \eq}$$
gives the couplings of $S$ to the NMSSM fields.
In addition we will have soft SUSY breaking terms of the form \cite{hm}
$${ (m_{s}^{2}\pm H^{2}) |\phi_{i}|^{2} 
+ (A_{\alpha} W_{\alpha} + h.c.)    \eq},$$
where $\phi_{i}$ are the scalar fields and $W_{\alpha}$ 
are the trilinear superpotential terms. 
($A_{\alpha} \sim m_{s} \pm H$ in the presence of 
non-zero $H$, with the sign of the $H$ 
correction being model dependent \cite{hm}).
With these superpotential and soft SUSY breaking terms, 
the $S$ scalar potential is assumed to be given by
$${ V(S) = (-m_{s}^{2} - H^{2}) |S|^{2} 
+ (A_{\lambda_{1}}\frac{\lambda_{1}}{3}S^{3}
+ h.c.) + 
(\lambda_{1}^{2} + \lambda_{2}^{2})|S|^{4}  \eq}.$$
The expectation value of $S$ is then given by
$${ <S> \approx \frac{(m_{s}^{2} + H^{2})^{1/2}}{(\lambda_{1}^{2} 
+ \lambda_{2}^{2})^{1/2}}   \eq}.$$
(We have neglected the $A_{\lambda_{1}}$ term, which can alter $<S>$
by at most factor of the order of 1).

       In this we have ignored the $N$ expectation value. 
In fact, a non zero $<N>$ will generally 
exist at the minimum together with $<S>$. This could result in a large 
effective $\mu$ term and affect the natural generation of the weak scale. 
We will discuss $N$ expectation value and its 
consequences later in this section. 

    We first consider the conditions under which 
there is no thermal restoration of the $Z_{3}$ symmetry.
This requires that the $S^{2}$ term in the finite 
temperature effective potential \cite{eu,dj} remains negative 
for all values of $T$ for which the 
$S$ condensate scalars are in thermal equilibrium. 

               The $N$ particles and Higgs 
particles $H_{i}$ will contribute to the 
$S$ finite-temperature effective potential if they are lighter than $T$. 
In fact, they will be lighter than $T$ for $T$ large compared with 
$T_{ew}$, since any $N$ or $H_{i}$ 
mass term due to $<S>$ cannot be very large 
compared with the weak scale without disrupting 
electroweak symmetry breaking \cite{ets}. 
The $S$ particles orthogonal to $<S>$, which we
 denote by $S^{'}$, will also gain a mass from $<S>$ and 
can contribute to the finite temperature effective potential.
The condition for these particles 
to be light compared with $T$ for $H \lae m_{s}$ is that
$${ \frac{\lambda_{1}}{(\lambda_{1}^{2} 
+ \lambda_{2}^{2})^{1/2}}  \lae \frac{T}{m_{s}}  \eq},$$
which will generally be satisfied for $T \gae T_{ew}$. 
(We need not consider the constraints for 
$H \gae m_{s}$, since in this 
case the soft SUSY breaking mass squared 
in the scalar potential becomes of the order of 
$H^{2}$, which will increase with $T$ much 
faster then the thermal $T^{2}$ correction, 
thus ensuring that the $Z_{3}$ symmetry 
remains broken at $H \gae m_{s}$ if it is broken for all $
H \lae m_{s}$). 
The effective SUSY $N$ mass from the $\lambda_{3}$ coupling is then given by
 $m_{N}(S) = 2 \lambda_{3} S$. Similar 
contributions will come from $\lambda_{1}$ and $\lambda_{2}$.

             We must then ensure that the temperature dependent 
contribution to the $S^{2}$ term in the finite temperature effective 
potential due to $m_{N}$, $m_{N}^{2}(S) T^{2}/8$, 
is small compared with $m_{s}^{2} S^{2}$ 
for all $ T \lae T_{eq}$, where $T_{eq}$
 is the temperature at which the $S$ scalars 
come into thermal equilibrium. 
The dominant process bringing the $S$ scalars into thermal equilibrium with
 the light particles in the thermal background will be inverse decays, 
with a rate $\Gamma_{inv} \approx 
\frac{\kappa_{d} \lambda_{i}^{2}}{4 \pi}T $ (where the factor $\kappa_{d}$ is 
not too small compared with 1; for the case of stop quarks this has been 
estimated to be typically of the order of 0.1 \cite{enqrot}).
If $T_{eq} \lae T_{R}$, then the thermal 
equilibrium condition $\Gamma_{inv} \gae H$ 
occurs during radiation domination and $T_{eq}$ is given by
$${T_{eq} \approx 
\frac{\kappa_{d} \lambda_{i}^{2} M_{Pl}}{4 \pi k_{T}}   \eq},$$
where the expansion rate during radiation domination is given by 
$H = \frac{k_{T}(T) T^{2}}{M_{Pl}}$, with 
$k_{T}(T) \approx 16$ for $T$ larger
 than around 100GeV \cite{eu}. 
This is consistent with $T_{eq} \lae T_{R}$ if 
$${  \lambda_{i} \lae 9 \times 10^{-6} 
\left( \frac{4 \pi k_{T}}{\kappa_{d}} \right)^{1/2}
 \left(\frac{T_{R}}{10^{9}GeV}\right)^{1/2} 
          \eq}.$$
If this is not satisfied then $T_{eq} \gae T_{R}$ and, with $H(T)$ given by 
$ H(T) = T^{4} ( k_{r}^{4}T_{R}^{2}M_{Pl})^{-1} $, the thermal equilibrium
temperature becomes
$${  T_{eq} \approx  \left(\frac{\kappa_{d}k_{r}^{4}}{4 \pi}\right)^{1/3} 
\lambda_{i}^{2/3} T_{R}^{2/3} M_{Pl}^{1/3}   \eq}.$$
The $Z_{3}$ symmetry remains broken at $T_{eq}$ if $\lambda_{i}
 \lae m_{s}/T_{eq} $. 
For $T_{eq} \lae T_{R}$ this requires that
$${   \lambda_{i} \lae \left(
\frac{4 \pi k_{T}}{\kappa_{d}}\right)^{1/3} 
\left(\frac{m_{s}}{M_{Pl}}\right)^{1/3} \approx 
1.3 \times10^{-5} \kappa_{d}^{-1/3} 
\left(\frac{m_{s}}{100GeV}\right)^{1/3}
 \eq},$$
where we have used $k_{T} \approx 16$.
This is the condition for avoiding 
$Z_{3}$ symmetry restoration if $T_{R} \gae 10^{7}GeV$.
If $T_{R}$ is less 
than $10^{7}GeV$, the upper limit on $\lambda_{i}$ occurs when $T_{eq} > 
T_{R}$ and the condition that 
the $Z_{3}$ symmetry remains broken at $T_{eq}$ becomes
$${ \lambda_{i} \lae 2.1 \times 10^{-6} \kappa_{d}^{-1/5}
\left(\frac{10^{9}GeV}{T_{R}}\right)^{2/5} 
\left(\frac{m_{s}}{100GeV}\right)^{3/5}
 \eq}.$$
For $ T_{R} = 10^{2}GeV \; (10^{5}GeV)$ 
this requires that $\lambda_{i} \lae 10^{-3} \;
(10^{-4})$. Thus we see that, for the case of 
$m_{s}^{2} < 0$ and renormalizible $S$ couplings to the NMSSM sector,
the couplings of $S$ to $N$ and to the Higgs fields must be rather small
if the reheating temperature is large compared 
with $T_{ew}$, in order not to 
thermally restore the $Z_{3}$ symmetry. 
In particular, if the reheating temperature is close to the 
gravitino upper limit, then these couplings 
must be less than around $10^{-5}$. 

                       We have not yet discussed the 
elimination of the domain walls.
 We will discuss this in more detail later, 
but for now we merely note that it is
very easy to introduce large $<S>$ dependent contributions into the $N$
 scalar potential. For example, from the superpotential 
couplings $k$ and $\lambda_{2}$, we obtain a term in the scalar potential
$${      - 2 \lambda_{2} k <S^{\dagger}>^{2} N^{2} \; + h.c.\;\; \sim  
 \left( \frac{2 \lambda_{2} }{\lambda_{1}^{2} + \lambda_{2}^{2}}\right) 
k m_{s}^{2} (N^{2} \; + h.c.)    \eq}.$$
However, if we assume that $k$ is not very small 
compared with 1, such that $2 k > \lambda_{2}$, 
then $\lambda_{2}$ 
must be smaller than $\lambda_{1}^{2}/k$ 
in order to avoid introducing a mass term for 
the $N$ scalars much larger than $10^{2}GeV$.
 In the NMSSM broken by radiative corrections, 
the phenomenologically favoured region of parameter space
has $<N> \gae 1TeV$. In this case the 
$N$ particles essentially decouple and one has an effective MSSM 
with $\mu \equiv \lambda_{N} <N>$ and a 
contribution to the soft SUSY breaking "$B$-term" given by 
$\nu = k <N>$ \cite{ets}. Both of these cannot be 
very much larger than $10^{2}GeV$ without disrupting the 
natural generation of the weak scale. 
(It is possible to have very large $<N>$, so long as 
$\lambda_{N}$ and $k$ are correspondingly small. The condition for 
a weak scale $\mu$ term is then that $\lambda_{N}/k \sim 1$ \cite{ets}). 
The effect of a large mass term from equation (13) would be to introduce an
effective negative mass squared, 
$m_{N\;eff}^{2}$, for the $N$ scalars, 
with the $N$ expectation being given by 
$<N> \approx |m_{N\;eff}|/k$, which would result 
in $\nu \approx |m_{N\;eff}|$ \cite{ets}.
 Requiring that $\nu$ be small compared 
with a mass scale $m_{*}$ not much larger than 
$10^{2}GeV$ and that the no-thermalization 
condition (12) is satisfied then implies that
$${ \lambda_{2} \lae 2.2 \times 10^{-10} \kappa_{d}^{-2/5}
\left(\frac{0.01}{k}\right)
\left(\frac{10^{9}GeV}{T_{R}}\right)^{4/5}
\left(\frac{m_{s}}{100GeV}\right)^{-4/5} 
\left(\frac{m_{*}}{100GeV}\right)^{2} 
\eq}.$$
(For $T_{R} \gae 10^{7}GeV$ the upper bound from (14) with 
$T_{R} \approx 10^{7}GeV$ applies). 
Thus for large reheating temperatures, 
$T_{R} \gae 10^{7}GeV$, we would require 
that $\lambda_{2} \lae 10^{-7} (0.01/k)$ 
in order to avoid a mass term for the $N$ scalars 
much larger than $10^{2}GeV$. For 
smaller $T_{R}$ this upper bound becomes weaker, 
allowing $\lambda_{2}$ to be as large as
$10^{-3}(0.01/k)$ for $T_{R} \approx 10^{2}GeV$. 
Therefore, unless $k \lae 10^{-4}$ 
(and, since $\lambda_{N}/k \sim 1$ is necessary \cite{ets},
 $\lambda_{N} \lae 10^{-4}$), for large $T_{R}$ there will
 be a tighter upper bound on some of the 
couplings than that coming from $Z_{3}$ symmetry restoration. 

                  We next consider the $N$ 
expectation value. When $<S>$ is introduced into the 
$N-S$ scalar potential, terms linear, 
quadratic and cubic in $N$ arise, which result in a
non-zero $<N>$. In general it is difficult to 
discuss the minimization of the potential analytically, 
but for the natural case where the couplings $\lambda_{i}$ 
(i=1,2,3) all have the same magnitude 
and $|k|$ is large compared with $|\lambda_{i}|$, 
and where $<N>$ does not alter the 
value of $<S>$ significantly, we find 
that $<N>$ is essentially determined by the linear and quartic terms in the
$N$ scalar potential. In this case $<N>$ is given by
$${ <N> \approx \left(\frac{\lambda_{i}}{k}\right)^{2/3} <S>     \eq}.$$  
(This will not affect the 
value of $<S>$ so long as $<N>$ is small compared with $<S>$). 
The main effect of $<N>$ will be to 
introduce an effective $\mu$ term for the Higgs doublets 
via the $\lambda_{N}$ superpotential coupling,
$${ \mu \approx \left(\frac{k}{\lambda_{i}}\right)^{1/3} 
\left(\frac{\lambda_{N}}{k}\right) m_{s}  \eq}.$$
Since $\lambda_{N}$ and $k$ must be of the same order of magnitude to 
naturally generate a weak scale $\mu$ term, we see that $\lambda_{i}$ 
cannot be too small compared with $k$, thus 
favouring large $\lambda_{i}$ and/or small $k$. 
We note that $<S>$ also introduces a $\mu$ term via the $\lambda_{SH}$ 
superpotential coupling. This will be acceptable so long 
as $\lambda_{SH}$ is not larger than 
$(\lambda_{1}^{2}+ \lambda_{2}^{2})^{1/2}$. 

                  We should check that spontaneous 
breaking of the discrete symmetry does not 
introduce dangerous quadratic divergences. 
A quadratically divergent tadpole
 arises at two-loops in $N=1$ supergravity with hidden sector 
SUSY breaking and no
 direct coupling of the hidden and observable 
sectors in the Kahler potential \cite{bag}. With a Kahler potential
given by
$${ K = N^{\dagger}N + H_{i}^{\dagger}H_{i} 
+ O\left(\frac{1}{M_{Pl}^{2}}\right)   \eq},$$
corresponding to minimal scalar kinetic terms,
and superpotential terms of the form \cite{bag}
$${ W = ... + \lambda NH_{1}H_{2} 
- \frac{\lambda_{\alpha}}{M_{Pl}}N^{2}H_{1}H_{2}
+ O\left(\frac{1}{M_{Pl}^{2}}\right)   \eq}$$
a divergent $N$ tadpole is generated
$${   \frac{2 \lambda \lambda_{\alpha} 
\Lambda^{2}}{(16 \pi^{2})^{2}}
\int d^{4} \theta e^{K/M_{Pl}^{2}}\left( 
\frac{N + N^{ \dagger } }{M_{Pl}} \right)   
\eq}.$$
On introducing the 
hidden sector SUSY breaking F-term for $K$,
$F_{K}^{2}\theta^{2}\overline{\theta}^{2}$ with $F_{K} = 
\left(\frac{m_{s}M_{Pl}}{ \sqrt{8 \pi}}\right)$, 
there is a contribution to the 
$N$ scalar potential of the form \cite{bag}
$${   \frac{2 \lambda \lambda_{\alpha} 
\Lambda^{2}}{(16 \pi^{2})^{2}M_{Pl}^{3}}
\left(\frac{m_{s}M_{Pl}}{ 
\sqrt{8 \pi}}\right)^{2}(N + N^{\dagger})   \eq}.$$
Taking the cut-off $\Lambda$ to be $M_{Pl}$, 
we find that the additional term in the 
$N$ scalar potential is of the form $ 
\overline{M}^{3}(N+N^{\dagger})$, where
$\overline{M} \approx 7 \times 10^{5}
\; \lambda^{1/3} \lambda_{\alpha}^{1/3}  GeV$. If
 $\overline{M}$ is much larger than $10^{2}GeV$, 
this term will make impossible the generation
 of a naturally weak scale $\mu$ term. 
Throughout this paper we will consider
 the natural value of non-renormalizible couplings such 
as $\lambda_{\alpha}$ to be of the
 order of 1, with the strength of the couplings set 
by the large mass scale which we will take
 to be the Planck scale. In this case we would 
require that $\lambda \lae 10^{-12}$
 in order to have $\overline{M} \lae 100GeV$. In the case 
where discrete symmetry breaking
 is spontaneous and non-vanishing at zero temperature, 
we can still have quadratically divergent tadpoles. 
For example, for $S$ with charge 
$Q(S) = 1/3$, one can have a non-renormalizible operator 
of the form of the $\lambda_{\alpha}$
 term but with an overall factor of $\left(\frac{<S>}{M_{Pl}}\right)^{2}$. 
For $S$ with charge $Q(S) =1/6$ this factor becomes
$\left(\frac{<S>}{M_{Pl}}\right)^{4}$. As a result, 
the quadratically divergence is 
not a problem for all $\lambda \lae 1$ if 
$<S> \lae 10^{13}GeV$ for the $Q(S) = 1/3$ case 
and $<S> \lae 10^{16}GeV$ for the $Q(S) = 1/6$ case. 
From equation (6) we see that, for 
$m_{s} \approx 10^{2}GeV$, $<S> \lae 10^{13}GeV$ 
requires that $(\lambda_{1}^{2} 
+ \lambda_{2}^{2})^{1/2} \gae 10^{-11}$. This can 
be easily satisfied.  
\newpage {\bf \underline {(b) $m_{S}^{2} > 0$}.  }
 
             In this case we will have a coherently 
oscillating $S$ scalar. We have to ensure that
 the $Z_{3}$-breaking condensate does not 
thermalize or decay before the electroweak 
phase transition. 

            The requirement that the condensate does not 
thermalize via inverse decays imposes the constraint, 
$${   \lambda_{i} \lae \left(\frac{4 \pi k_{T}}{\kappa_{d}}\right)^{1/2} 
\left( \frac{T}{M_{Pl}} \right)^{1/2}    \eq},$$
where we have assumed that the Universe is radiation dominated. 
Thus, with $ T \approx T_{ew}$, 
this requires that $\lambda_{i} \lae 10^{-8}$. 
(Below $T_{ew}$ the $S^{'}$, $H_{i}$ and $N$ particles gain
masses from electroweak symmetry breaking).
This constraint should be applied if 
the masses of the $S^{'}$, $H_{i}$ and $N$ particles due to $<S>$ are
small compared with T. During the inflaton oscillation 
dominated era, the energy density of the $S$
oscillations scales as that in the inflaton oscillations,
whilst after reheating the energy density of $S$ oscillations scales
as $ a(T)^{-3}$ , where $a(T)$ is the scale factor. 
Thus the energy density during radiation domination at $T < T_{R}$ is given by
$${ \rho_{S} (T) =  \frac{\alpha_{\rho}(T)
T^{3} T_{R}}{m_{s}^{2} M_{Pl}^{2}} \rho_{o} \eq}, $$
where $\rho_{o}$ is the initial energy density of the $S$ condensate and 
$\alpha_{\rho}(T) = \frac{g(T)}{g(T_{R})} 
k_{T}\left(T_{R}\right)^{2}$. The  
the corresponding amplitude of the $S$ 
oscillations at temperature $T$ is 
$ S(T) \approx  \left( \rho_{S}/\rho_{o} \right)^{1/2} S_{o}$,
 where $S_{o}$ is the initial amplitude of the oscillations.
The condition that the mass of the particles
 due to their interaction with the condensate \cite{ad}, 
$\lambda_{i} S(T)$, is less than $T$ then becomes
$${ \lambda_{i} \lae \left( \frac{\left( \lambda_{1}^{2} 
+ \lambda_{2}^{2}\right)}{\alpha_{\rho}(T)}
 \frac{M_{Pl}^{2}}{T T_{R}} \right)^{1/2}    \eq}.$$
This will be easily satisfied 
for $T_{ew} \lae T \lae T_{R}$. Therefore (21) applies. 
 We should also check that 
the energy density is not dominated by the coherent $S$ oscillations. This is 
satisfied if $\lambda_{1}^{2} + \lambda_{2}^{2} \gae 10^{-28}m_{s}^{2}/T$, 
which is typically easily satisfied. 

             Under the assumption that the 
$S$ condensate particles are heavy enough to
decay directly to a two-body final state via the coupling 
$\lambda_{i}$, with a rate $\Gamma_{d} \approx 
\frac{\lambda_{i}^{2}}{4 \pi}m_{s}$, the requirement 
that the condensate does not decay before temperature 
$T$ imposes the constraint
$${ \lambda_{i} \lae \left(4 \pi k_{T}\right)^{1/2} 
\left( \frac{T^{2}}{m_{s} M_{Pl}} \right)^{1/2}
    \eq}.$$
This will give an even tighter constraint on $\lambda_{i}$ than the 
no-thermalization constraint (21) if the coherent
 oscillations have to survive down to a 
temperature much lower than the electroweak
phase transition in order to eliminate the 
domain walls, as is likely to be 
the case in the NMSSM \cite{saw}. 

               We next consider the conditions under which the $S$
condensate can eliminate the NMSSM domain walls. This
requires that the pressure due to the energy difference
$\Delta V$ between the domains is large enough to
cause the higher energy domains to collapse on the time
scale $H(T)^{-1}$ \cite{dwp,dwp2,dwp3}. The 
condition on $\Delta V$ depends on
whether the domain walls are relativistic or not, i.e. 
whether there are significant frictional forces due to the interaction
of the domain walls with the thermal background particles. 
If there are strong frictional forces, then the walls will expand 
until the force due to surface tension is balanced by the
 friction, which will occur at a radius 
much less than the horizon $H^{-1}(T)$. 
The pressure due to $\Delta V$ must then 
overcome this large frictional force. 
If there are effectively no frictional 
forces, meaning that the surface tension 
can overcome the frictional forces for 
relativistic walls of radius equal to the horizon, 
then the condition on $\Delta V$ will be typically much weaker. 
For the case of an unbroken $Z_{3}$ symmetry, the conventional 
view would be that there
are no (or at least highly suppressed) 
frictional forces, since the particles crossing the 
domain walls would have equal masses on either side of the wall 
and so would transfer no net momentum to the wall on passing through 
\cite{dwp3,dwp4}. 
However, it has been suggested that, 
on taking into account the change in the particle masses in the 
vicinity of the wall, there will be a non-zero 
reflection coefficient, leading to a large 
 frictional force until the 
heavier particles become Boltzmann 
suppressed at temperatures less than around that of the
 quark-hadron phase
transition, after which the domain walls become 
relativistic on the scale of the horizon 
\cite{saw}. 
In more detail, the friction force 
per unit area on the wall is estimated to be $f \approx 
\frac{g}{8 \pi^{2}} \frac{m^{5}}{m_{W}^{2}} v T$, 
where $m$ is the mass of the 
heaviest thermal background particle with
$m < T$, $g$ is the number of degrees of freedom
of the particle  and $v$ is the velocity of the wall \cite{saw}. 
This force rapidly decreases as $T$ decreases and
 the heavier thermal particles become Boltzmann suppressed. 
The condition on $\Delta V$ to
 eliminate the domain walls, $\Delta V \gae f$, will
 be most easily satisfied once the frictional force is weak enough
 to allow the domain walls to remain
 relativistic up to scales of the order of the horizon. 
Thereafter, the condition will become more difficult
 to satisfy as $T$ decreases, since the pressure difference between the 
domains rapidly decreases as $<S>$ decreases with $T$.

              Suppose the domain walls become relativistic
 on horizon scales at $T_{rel}$.
 Soon after $T_{rel}$, the uncollapsed domain walls
will have a radius of the order of the horizon 
$H(T_{rel})^{-1}$. This will continue to be
true until the pressure due to $\Delta V$ dominates the force due to the 
surface tension $\sigma$. This in general occurs once the radius of a 
domain is larger than a critical radius, $ r_{c}$, given by
$${ r_{c} \approx \frac{\sigma}{|\Delta V|}   \eq}.$$ 
The smallest $|\Delta V|$ will therefore 
correspond to the largest possible $r_{c}$, the horizon radius.
Although the $Z_{3}$ domain wall will consist of the Standard Model Higgs
 fields as well as the $N$ field, for simplicity we will consider the
 wall to be made only of the $N$ field. 
In this case the domain wall potential 
will be due to the soft SUSY breaking $N^{3}$ 
A-term and the surface tension will be $\sigma \approx m_{s} <N>^{2}$
\cite{dwp2,dwp3}.

              From the superpotential couplings 
$\lambda_{2}$ and $k$ we obtain an energy density
splitting between the different $Z_{3}$ vacuum states 
$${ \Delta V \approx 2 \lambda_{2} k S^{2} N^{2}   \eq}.$$
This depends on the value of $S^{2}$ and so will have a non-zero value on 
averaging over the $S$ oscillations.
The condition that this can overcome the surface tension for horizon-sized
 relativistic domains at $T_{rel}$ is then that
$${ \frac{\lambda_{1}^{2} + \lambda_{2}^{2}}{2 k \lambda_{2}} 
\lae \frac{\alpha_{\rho}(T_{rel})}{k_{T}(T_{rel})}
\frac{T_{R} T_{rel}}{m_{s} M_{Pl}}    \eq}.$$
The left-hand side will be smallest for 
$\lambda_{1} \lae \lambda_{2}$, so we require that 
$${\frac{\lambda_{1}}{k} \lae \frac{\lambda_{2}}{k} 
\lae  \frac{10^{-13} \alpha_{\rho}(T_{rel})}{k_{T}(T_{rel})}
\left(\frac{T_{R}}{10^{9}GeV}\right)  \left(\frac{T_{rel}}{0.1GeV}\right)  
\left(\frac{100 GeV}{m_{s}}\right)
\eq}.$$ 
Therefore we see that the requirement that $\Delta V$ is large enough
 to eliminate the domain walls imposes
a very tight constraint on at least some of the couplings. 
A similar $S^{2}$ dependent contribution to
 $\Delta V$, $\Delta V \approx 2 \lambda_{1} \lambda_{3} S^{2} N^{2}$,
 arises from the superpotential couplings 
$\lambda_{1}$ and $\lambda_{3}$. For this to
eliminate the domain walls we would require that
$${\frac{\lambda_{2}}{\lambda_{3}} \lae \frac{\lambda_{1}}{\lambda_{3}}
 \lae  \frac{10^{-13} \alpha_{\rho}(T_{rel})}{k_{T}(T_{rel})}
\left(\frac{T_{R}}{10^{9}GeV}\right)  \left(\frac{T_{rel}}{0.1GeV}\right)  
\left(\frac{100 GeV}{m_{s}}\right)
\eq},$$ 
which would demand even smaller values of 
$\lambda_{1}$ and $\lambda_{2}$ once the 
no-thermalization and decay constraints, (21) and (24), are imposed. 

                 Thus we can conclude that, for the case 
of renormalizible couplings of the $Z_{3}$ symmetry breaking
scalar $S$ to the NMSSM, very small couplings 
(less than $10^{-8}$ and in some cases less than $10^{-13}$) 
are required in order to eliminate the NMSSM domain walls 
if the hidden sector mass squared of $S$ is
positive. 
If it is negative, small couplings 
$\lae 10^{-5}$ (and even smaller if the NMSSM couplings
 $k$ and $\lambda_{N}$ are larger than $10^{-4}$) are again 
required if the reheating temperature is 
larger than $10^{7}GeV$, but, for smaller
values of $T_{R}$, the constraints from 
avoiding thermal restoration of the $Z_{3}$ symmetry 
become somewhat weaker, allowing 
couplings as large as $10^{-3}$ for $T_{R} \approx T_{ew}$. 

\newpage {\bf \sec. Non-Renormalizible $S$ couplings to the NMSSM Sector}. 

             We have seen that the case of renormalizible 
$S$ couplings to the NMSSM sector 
requires rather small couplings when the reheating temperature is large 
(less than $10^{-5}$ for the case where 
the reheating temperature 
is larger than $10^{7}GeV$, with an even smaller upper bound 
if the NMSSM $N$ couplings are not less than about $10^{-4}$). 
In particular, for the case of a positive hidden sector mass
 squared for the $S$ scalar, some couplings must be 
less than about $10^{-13}$, regardless of the reheating temperature.
 One way to overcome the 
need for such small couplings might be to simply 
eliminate $all$ the renormalizible 
couplings of the $S$ scalar to the NMSSM fields, 
allowing only non-renormalizible couplings 
suppressed by powers of some large mass scale 
(which we will take to be the Planck scale). 
This is naturally achieved by extending the $Z_{3}$ symmetry
 of the NMSSM to a discrete symmetry, which we will call $Z_{A}$,
under which the charge of $S$ is such that no 
renormalizible couplings of $S$ to the NMSSM
 fields are possible. 
\newline {\bf \underline{(a) $m_ {s}^{2} < 0$}.}

                In this case there will be a 
large and constant expectation value for the $S$ scalar once 
$H \lae m_{s}$. This large expectation 
value will introduce what we believe to be a 
generic problem in this case, namely
 a very large SUSY mass for $N$, much larger than the weak scale.
 To see this, consider the simplest example, corresponding to $S$ having a 
$Z_{A}$ charge 1/6 whilst $N$, as usual, has charge $1/3$. Then
$S$ can only have leading order non-renormalizible 
superpotential couplings of the form
$${ \frac{\lambda}{M_{Pl}} S^{2}N^{2} + 
\frac{\lambda^{'}}{M_{Pl}} S^{2}H_{1}H_{2}
+ \frac{\lambda_{S}}{6 M_{Pl}^{3}} S^{6}  \eq}.$$
 The expectation value following from the $S$ scalar potential, 
$${  V(S) = -m_{s}^{2} S^{2} 
+ \left(\frac{\lambda_{S}}{M_{Pl}^{3}}\right)^{2} 
S^{10}    \eq},$$
gives, with $m_{s} \approx 10^{2}GeV$ and 
$\lambda_{S} \approx 1$, $<S> \approx \rm 5\times 10^{14}GeV$. 
This results in an SUSY mass for $N$ 
of the order of $10^{10}GeV$, making a solution of the
 $\mu$ problem impossible in this case. 
We find that decreasing the $Z_{A}$ charge 
of $S$ or introducing an intermediate
mass scale (in the manner of the 
model to be described in 
the following section) only 
increases the $N$ mass. Based on this, we conclude that the
 case of $S$ scalars with 
non-renormalizible couplings and a negative 
hidden sector mass squared is unlikely to 
provide a natural solution of the NMSSM domain wall
problem.
\newline {\bf \underline{(b) $m_ {s}^{2} > 0$} }.

                 So far we have found 
no model which can solve the NMSSM domain wall problem for 
reheating temperatures much larger 
than the electroweak phase transition temperature
 without requiring small couplings. 
We now present a model, based on a coherently oscillating scalar, 
which can eliminate the domain wall problem of the 
NMSSM for values of $T_{R}$ as large as the gravitino upper bound 
and which does not require any small couplings. 

               The model we consider requires, in addition to the
NMSSM singlet $N$ and the 
$Z_{A}$ breaking field $S$, two additional singlet fields, 
$X$ and $B$. $X$ will acquire an 
intermediate mass as a result of an expectation value for $B$.
 The model is also assumed to have a
more complicated discrete symmetry, $Z_{A}\times Z_{B}$, 
where $Z_{A}$ is the
extension of the original NMSSM $Z_{3}$ whilst $Z_{B}$ is a
discrete symmetry under which all the NMSSM fields are singlets.
 $Z_{B}$ is introduced in order to control the non-renormalizible terms. 
The discrete symmetry charges of the fields are as defined in Table 1(a). 
\newline
{\bf Table 1(a). Field charges under $Z_{A}\times Z_{B}$.}  

\begin{center}
\begin{tabular}{|c||c|c|c|c|}          \hline
& $S$ & $X$ & $B$ & $N$ \\ \hline
$Z_{A}$ & $\frac{1}{6}$ &  $\frac{1}{2}$ &  0 & $\frac{1}{3}$
 \\
$Z_{B}$ & -$\frac{1}{3}$ &  $\frac{1}{3}$ &  $\frac{1}{6}$ & 0
 \\
 \hline
\end{tabular}
\end{center} 

It may well be significant that the $Z_{A}\times Z_{B}$ 
charges of the remaining NMSSM fields can always be chosen to
eliminate the dangerous renormalizible $B$ and $L$ violating operators 
$u^{c}d^{c}d^{c}$, $d^{c}QL$ and $e^{c}LL$ \cite{nilles}.
For example, this is true for the charges of Table 1(b).
 Therefore the absence of these renormalizible interactions 
could be interpreted as a sign of the existence of a
non-trivial discrete symmetry such as $Z_{A} \times Z_{B}$, rather than 
of a simple R-parity or conservation of B and L. 
\newpage
{\bf Table 1(b). 
Example of field charges eliminating dangerous renormalizible operators.}

\begin{center}
\begin{tabular}{|c||c|c|c|c|c|c|c|}          \hline
& $e^{c}$ & $L$ & $d^{c}$ & $H_{d}$ & $Q$ & $u^{c}$ & 
$H_{u}$
\\ \hline
$Z_{A}$ &  -$\frac{1}{6}$  &  $\frac{1}{2}$  
&  $\frac{1}{3}$  &  -$\frac{1}{3}$  &  0  &  0  &  0 
\\
$Z_{B}$ &  $\frac{1}{3}$  &  -$\frac{1}{3}$  &  0 &  0  &  0  & 0 & 0  
\\
\hline
\end{tabular}
\end{center} 
The superpotential of the model is then given by 
$${ W = W_{NMSSM} + W_{X} + W_{NR} \eq},$$ 
where $W_{X}$ is the allowed renormalizible coupling of $X$, 
$${ W_{X} = \lambda_{a} XSN      \eq}$$
and $W_{NR}$ gives 
the leading order non-renormalizible superpotential couplings involving the
 additional gauge singlet scalars, 
 $${ W_{NR} =  
\frac{\lambda_{S}}{6 M_{Pl}^{3}} S^{6} + 
\frac{\lambda_{BX}}{M_{Pl}} B^{2} X^{2} +
\frac{\lambda_{B}}{6 M_{Pl}^{3}} B^{6} \eq}.$$ 
The model will require an intermediate mass for the $X$ scalar. 
This will be generated by an 
expectation value for $B$. The obvious 
possibility is to assume that the hidden sector mass squared 
of the $B$ scalar is negative, 
which would result in an expectation value for $B$ given by, 
$${ <B>  \approx
\left(\frac{m_{s}^{2} M_{Pl}^{6}}{5 \lambda_{B}^{2}}\right)^{1/8} 
\approx 5 \times 10^{14}GeV   
\eq}.$$ 
However, we should perhaps note that in this case there could be 
problem due to the late decay of oscillations 
of the weakly coupled $B$ scalar around the
 minimum of its potential.
 The $B$ scalar has a mass of the order of $m_{s}$,
 whilst the $X$ particles to which it couples gain a mass 
of the order of $10^{10}GeV$ as a result of $<B>$. 
Thus the $B$ decay rate will be highly
 suppressed in this case. 
If the $B$ field is not 
very close to 
the minimum of its potential at the end of inflation, 
its coherent
oscillations about the minimum of its potential will decay long after
nucleosynthesis and could dominate the energy density of the Universe
when they decay. This is all dependent upon
 the value of $B$ at the end of inflation
and the dynamics of 
its subsequent evolution as the Universe expands 
and its effective (negative) $O(H^{2})$
SUSY breaking mass term decreases.  
Since in our model the $B$ field only serves to generate an 
intermediate mass for $X$ and, in a sense, 
may be regarded as a toy field representing
the dynamics of a more realistic model, 
we will not concern ourselves 
with the details of this issue here.

                   In general we will consider
$M_{X} \gae 10^{10}GeV$. One reason for 
this choice, as we will show later, 
is that it ensures that the $X$ particles 
are Boltzmann suppressed
 for all temperatures, thus eliminating 
the possibility of thermalization
 of the $S$ condensate by light thermal
$X$ particles. Integrating out the massive X field, we obtain 
the following effective interaction in the superpotential
$${\frac{\lambda_{a}^{2}}{M_{X}} S^{2} N^{2}  \eq}.$$ 
This is valid so long as $ \lambda_{a} <S> \lae M_{X}$, such that the 
mixing between the $X$ and $N$ fields via the coupling (33) may be neglected.
This interaction is fundamentally important in what follows, as it 
will allow both a sufficiently rapid decay of the $S$
oscillations so as to avoid problems with nucleosynthesis and 
at the same time allow a sufficiently strong interaction of the $S$
oscillations with the $N$ scalars to drive away the $Z_{3}$ domain
walls. Assuming that the $S$ scalar mass is large enough, 
the $S$ scalars will decay to $N$ particles and $S$ fermions 
(which will be highly decoupled and will 
typically have a mass not much larger
than 1eV, coming from the interaction (36) with $<N> \sim 100GeV-1TeV$). 
 So long as their energy density is sufficiently small compared with
than that of the radiation when they are produced, 
 the $S$ fermions will not give rise to any
 cosmological problems. (This may be difficult to achieve, however, if 
the Universe is dominated by the $S$ scalars when they decay, since one 
would expect approximately 1/3 of the energy density from $S$ decays to be in
the form of $S$ fermions. 
Alternatively, for the charge assignment of 
Table 1(b), one can obtain an operator of the same
 form as equation (36) but with $S^{2}N^{2} \rightarrow SNH_{u}L$,
which can allow the $S$ scalars to safely 
decay to NMSSM fields). 

            The $S$ scalar potential is given by
$${V(S)
\approx (m_{s}^{2} - H^{2}) S^{2} + \left(
\frac{\lambda_{S}}{M^{3}} S^{5} \right)^{2}    \eq}.$$ 
The initial value of the $S$ field when 
the oscillations begin at $H \approx m_{s}$, $S_{o}$, 
is therefore given by
$${ S_{o}
\approx \left(\frac{m_{s}^{2} M_{Pl}^{6}}{5
\lambda_{S}^{2}}\right)^{1/8}   \eq}.$$
We first consider the conditions
under which the $S$ energy density decays at a temperature
less than that of electroweak phase transition, $T_{ew}$, without
disturbing the predictions of nucleosynthesis.
 This requires either that the ratio of the energy density in
the $S$ scalars to that in radiation when the $S$ scalars 
decay, $r_{D}$, is less than $10^{-6}$, in order not to photodissociate 
 the helium abundance \cite{gravb2,helium}, 
or that the $S$
scalars decay at a temperature greater than around 10MeV \cite{hadron}.
The former condition requires that 
$${ T_{R} \lae \frac{3
r_{D}}{8 \pi} \frac{M_{Pl}^{2}}{S_{o}^{2}} T_{d} 
  \eq}.$$
The $S$ scalars in the condensate decay via the interaction of
equation (36) with a rate given approximately by
$${ \Gamma_{d} \approx \frac{\alpha_{d} 
\lambda_{a}^{4} m_{S}^{3}}{M_{X}^{2}}      \eq},$$
where $\alpha_{d} \approx \frac{1}{192 \pi^{3}} \approx 10^{-4}$
 and we are assuming that 
the $S$ scalars are heavy compared with the $N$ particles.
 This gives for the decay temperature, 
assuming that decay occurs during radiation domination, 
 $${ T_{d} \approx
\frac{\alpha_{d}^{1/2} \lambda_{a}^{2}}{k_{T}(T_{d})^{1/2}}
\left(\frac{M_{Pl} m_{s}^{3}}{M_{X}^{2}}\right)^{1/2}   \eq}.$$ 
Therefore, in the case where decay occurs 
after nucleosynthesis at $T_{d} \lae 1 MeV$, 
we require from equation (39) that the
 reheating temperature satisfies, 
$${ T_{R} \lae r_{D} m_{s} \gamma \lambda_{a}^{2} 
\left(\frac{M_{Pl}}{M_{X}}\right)    \eq},$$
 where 
$${\gamma
= \frac{3 \alpha_{d}^{1/2} \lambda_{s}^{1/2}}{8 \pi 
k_{T}(T_{d})^{1/2}}   }.$$
Thus with $r_{D} \approx 10^{-6}$, $\gamma \approx 10^{-4}$, 
$m_{S} \approx 10^{2}GeV$ and $M_{X} \gae 10^{10}GeV$, this requires that 
$T_{R} \lae \alpha_{d}^{1/2} \lambda_{a}^{2} 10^{4}GeV$. 
This imposes a 
severe restriction on $T_{R}$ for 
$\lambda_{a} \lae 0.1$.  
On the other hand, from equation (41), 
we see that it is quite likely that $T_{d} \gae 10MeV$.
 In general, $T_{d}$ is given by
$${ T_{d} \approx 
10 \left(\frac{\alpha_{d}}{10^{-4}}\right)^{1/2}
\left(\frac{\lambda_{a}}{0.1}\right)^{2}
\left(\frac{m_{s}}{100GeV}\right)^{3/2}
\left(\frac{10^{10}GeV}{M_{X}}\right) MeV
\eq}.$$
 Therefore, in this case we expect that, 
for $M_{X} \approx 10^{10}GeV$ (which will be seen to 
be the preferred 
order of magnitude for $M_{X}$ in this model), the $S$ oscillations
will typically decay at a temperature between 10MeV and 1GeV for 
$\lambda_{a}$ in the range 0.1 to 1. 
The amplitude of the $S$ oscillations at $T < T_{R}$ 
(assuming that the Universe is radiation dominated) is given by
$${ <S> \approx \frac{16 \alpha_{\rho}(T)^{1/2}}{\lambda_{S}^{1/4}} 
\left(\frac{m_{s}}{100GeV}\right)^{3/4}
\left(\frac{T}{m_{s}}\right)^{3/2}
\left(\frac{T_{R}}{10^{9}GeV}\right)^{1/2} GeV   \eq}.$$
The corresponding energy density in the coherent $S$ oscillations is
$\rho_{S} \approx m_{s}^{2}S^{2}$. This will come to dominate the radiation 
energy density once $T \lae T_{dom}$, where
$${ T_{dom} \approx \frac{21}{\lambda_{s}^{1/2}}
\left(\frac{m_{s}}{100 GeV}\right)^{1/2}
\left(\frac{T_{R}}{10^{9}GeV}\right) GeV
\eq}.$$
Thus if the domain walls become relativistic below $T_{dom}$, the elimination of the 
domain walls by the condensate will occur whilst the Universe is dominated by the energy
density of the oscillating $S$ field and possibly 
whilst the background radiation is dominated 
by radiation coming from the decays of the condensate scalars. In the
 case where the $S$ scalars decay during matter
 domination, it would seem that equation (41), derived assuming 
radiation domination, is not correct. However, since the
 decay condition is given by $\Gamma_{d} \approx H \propto \rho_{S}^{1/2}$ 
and since the radiation energy of the Universe when all the $S$ scalars
decay will simply be given by $\rho_{S}$ at this
 time, the decay temperature will be the same as that given in equation (41).  

                    We must also ensure that the $S$
condensate is not thermalized by scattering
processes involving the thermal background particles. 
We first show that the $X$ particles are always heavier than $T$ 
and so Boltzmann suppressed if the coupling $\lambda_{a}$ is not
 too small compared with 1. 
The mass of the $X$ particles from the $X-N$ mass matrix
is given by $\overline{M}_{X} \approx Max(M_{X}, \lambda_{a}<S>)$. Therefore 
$\overline{M}_{X} \gae 10^{10}GeV$ if $M_{X} \gae 10^{10}GeV$. 
Since, from equation (1), $T_{r}$ at $H \approx m_{s}$ is given by 
$${ T_{r}(H\approx m_{s}) \approx 3.2 \times 10^{9} 
\left(\frac{T_{R}}{10^{9}GeV}\right)^{1/2} GeV  \eq},$$
we see that $\overline{M}_{X} \gae T$ for 
all $H$ less than $m_{s}$. At $H \approx m_{s}$
the $X$ mass will be given by 
$\overline{M}_{X} \approx \lambda_{a}<S>$ so long as 
$\lambda_{a} \gae 2 \times 10^{-5} \lambda_{S}^{1/4}
\left(\frac{M_{X}}{10^{10}GeV}\right)$. Assuming that this 
is satisfied, we see that since $<S> 
\propto H \propto T^{4}$ for $H \gae m_{s}$,  
$\overline{M}_{X}$ will be larger than $T$ 
for all temperatures after the end of inflation. 
Thus the $X$ particles will not thermalize the 
condensate for $\lambda_{a} \gae 10^{-5}$. 

      We next consider whether the $N$ 
particles can thermalize the condensate. 
The $N$ particle mass from the $X-N$ mass martix is given by 
$\overline{M}_{N} \approx Min(\lambda_{a}<S>, 
\frac{\lambda_{a}^{2} <S>^{2}}{M_{X}})$. In the following, 
for simplicity, we will
 concentrate on the case of large $T_{R}$, not 
much smaller than the gravitino upper bound, 
which is of most interest to us here.
 Smaller values of $T_{R}$ can be analysed in a 
similar way. Let $T_{N}$ be the temperature
 below which $\overline{M}_{N}$ becomes less than $T$. 
Let $T_{c}$ be the temperature below which 
$ \lambda_{a}<S> \lae M_{X}$. 
Since for
 $T \gae T_{c}$ the $N$ and $X$ particles 
have the same mass and we have shown than 
$\overline{M}_{X} \gae T$ for all $T$, 
$\overline{M}_{N}$ can only become less than $T$ at 
temperatures less than or equal to $T_{c}$. 
Assuming that $T_{c} \lae T_{R}$, $T_{c}$ is 
given by 
$${  T_{c} \approx  \frac{2 \times 10^{7}\;
\lambda_{s}^{1/6}}{\lambda_{a}^{2/3}} 
\left(\frac{m_{s}}{100GeV}\right)^{1/2} 
\left(\frac{M_{X}}{10^{10}GeV}\right)^{2/3}
\left(\frac{10^{9}GeV}{T_{R}}\right)^{1/3}GeV  \eq}.$$
Therefore with $\lambda_{a}$ not too small, 
say in the range 0.01 to 1, this will be less than
 $T_{R}$ for $M_{X} \approx 10^{10}GeV$ and 
for reheating temperatures not much 
smaller than the gravitino upper bound. 
$\overline{M}_{N}$ can only be less than $T$ 
if $T_{N} \lae T_{c}$. In this case $T_{N}$ is given by
$${T_{N} \approx  \frac{1.2 \times 10^{5} \lambda_{S}^{1/4}}{\lambda_{a}}
\left(\frac{m_{s}}{100GeV}\right)^{3/4}
 \left(\frac{M_{X}}{T_{R}}\right)^{1/2}
 GeV    \eq}.$$
This is indeed less than $T_{c}$ if 
$${\lambda_{a} \gae 7 \times 10^{-6} 
\lambda_{S}^{1/4} \left(\frac{m_{s}}{100GeV}\right)^{3/4}
\left(\frac{10^{10}GeV}{M_{X}}\right)^{1/2}
\left(\frac{10^{9}GeV}{T_{R}}\right)^{1/2}
    \eq}.$$
This will be easily satisfied for values of 
$T_{R}$ not much smaller than the gravitino upper bound, 
$M_{X} \approx 10^{10}GeV$ 
 and $\lambda_{a}$ in the range 0.01 to 1. 
Therefore in this case we must apply the no thermalization
 condition during radiation domination at $T_{N}$. 
Thermalization 
will primarily occur via the effective interaction of equation (36). 
We treat the scattering of the thermal $N$ particles from the 
zero momentum $S$ condensate scalars as a simple scattering
 process involving $S$ and $N$ particles.
The scattering
 rate for $S
\psi_{N} \leftrightarrow S \psi_{N}$ is then given by 
$${\Gamma_{sc} \approx \frac{1}{\pi^{3}}
\frac{\lambda_{a}^{4} T^{3}}{M_{X}^{2}}                  \eq}.$$
 Requiring that the no
 thermalization condition, $\Gamma_{sc} \lae H$, be satisfied 
at $T_{N}$, equation (48), then gives 
 for the no thermalization constraint on $T_{R}$,
$${  T_{R} \gae \; 6 \times 10^{12} \lambda_{a}^{6}
\left(\frac{m_{s}}{100GeV}\right)^{3/2} 
\left(\frac{10^{10}GeV}{M_{X}}\right)^{3}GeV   \eq}.$$
$\lambda_{a} \lae 0.2$ 
and $M_{X} \approx 10^{10}GeV$ will 
easily allow a range of values for $T_{R}$ below 
the gravitino upper bound to be consistent with 
no thermalization of the $S$ condensate. Thus with, for example,
$\lambda_{a} \approx 0.1$, $M_{X} \approx 10^{10}GeV$ and 
$T_{R} \approx 10^{7-9}GeV$, the $S$ condensate will evade
 thermalization and will safely decay at a temperature of around $10MeV$.

               We next consider the conditions under which the $S$
condensate can eliminate the NMSSM domain walls.
 We first consider the case where the
Universe is radiation dominated.
 The terms
in the SUSY scalar potential responsible for the energy density
splitting are given by
$${ \frac{2 k \lambda_{a}^{2}}{M_{X}} (S^{2}N
N^{\dagger \; 2} + h.c.) \eq}.$$
 This is explicitly dependent on
the phase of the $S$ field and has a 
non-zero average over time. It results in a splitting of 
energy density of the different $N$ vacuum states, $\Delta V$, given by
$${\Delta V \approx \frac{k \lambda_{a}^{2}}{M_{X}} <S^{2}>
<N>^{3}    \eq}.$$ 
Thus the condition for $\Delta V$ to be large 
enough to eliminate relativistic domain walls 
at a temperature $T$ during radiation 
domination, $\Delta V (T) \gae \sigma H(T)$, becomes
 $${ <S^{2}> \gae \frac{k_{T}(T)
T^{2}}{k \lambda_{a}^{2}} \frac{M_{X}}{M_{Pl}}\frac{m_{s}}{<N>}    \eq}.$$ 
Since $<S^{2}>$ is proportional to $T^{3}$, we see
that this is most easily satisfied as soon as the 
walls become relativistic at 
$T_{rel}$. Thus in order to eliminate the 
domain walls during radiation domination 
we require that 
$${ T_{R} \gae \frac{k(T_{rel})}{\alpha_{\rho}(T_{rel}) 
k \lambda_{a}^{2}} \frac{1}{T_{rel}} 
\frac{m_{s}}{<N>}
\frac{ m_{s}^{2} M_{X}
M_{Pl}}{S_{o}^{2}}   \eq}.$$ 
For example, if we consider the domain wall to 
become relativistic shortly after the 
quark-hadron phase transition \cite{saw}, say at $T_{rel}
 \approx 0.1GeV$, and to be too strongly 
damped for the pressure due to $\Delta V$ 
to be able to collapse the domain walls at 
higher temperatures, then we obtain
$${T_{R} \gae 5 \times 10^{3}
\frac{f_{c} \lambda_{s}^{1/2}}{k \lambda_{a}^{2}}
\left(\frac{m_{s}}{<N>}\right)
\left(\frac{m_{s}}{100GeV}\right)^{3/2}
\left(\frac{0.1GeV}{T_{rel}}\right)
\left(\frac{M_{X}}{10^{10}GeV}\right)
 GeV   \eq}.$$ 
We have included a correction factor, $f_{c}$, for the case 
where the domain walls 
become relativistic during $S$ condensate 
matter domination rather than during radiation
 domination. We have shown that the Universe becomes matter dominated 
at a temperature $T_{dom}$, equation (45), 
which is in the tens of GeVs for $T_{R}$ close to
 the gravitino upper bound. In this case, for $T_{rel} < T_{dom}$, we have
 to modify the domain wall constraint on $T_{R}$. 
Once $T \lae T_{dom}$, entropy will be effectively conserved and 
the radiation energy density will be mostly due to the primordial radiaiton. 
However, there will also be additional 
radiation due to the decay of the $S$ condensate,
 with a temperature given by 
equation (1) with the replacement $T_{R} \rightarrow T_{d}$. 
This radiation will dominate the 
primordial radiation once $T \lae T_{*}$, where
$${ T_{*} = \left(\frac{4 g(T_{d})}{25} \right)^{1/5} T_{d}^{4/5} 
T_{dom}^{1/5}    \eq}.$$
Once $T \lae T_{*}$ entropy 
is no longer effectively conserved and $H \propto T^{4}$
 drops much more rapidly with $T$. The result of all this
 on the domain wall bound is to 
introduce a correction factor $f_{c}$, where, for $T_{rel} > T_{*}$, 
$${ f_{c} = \left(\frac{T_{dom}}{T_{rel}}\right)^{1/2}    \eq}$$
and for $T_{rel} < T_{*}$
$${ f_{c} = \left(\frac{g(T_{*})}{g(T_{rel})}\right)^{1/2}
\left(\frac{T_{*}}{T_{rel}}\right)^{5}
\left(\frac{T_{dom}}{T_{*}}\right)^{1/2}
\eq}.$$
The weakest lower bound on $T_{R}$
 is obtained when $T_{rel} \gae T_{*}$. 
$T_{rel} \gae T_{*}$ is satisfied if 
$${ T_{d} \lae 40 \lambda_{s}^{1/8}
\left(\frac{1}{g(T_{d})}\right)^{1/4}
\left(\frac{T_{rel}}{0.1GeV}\right)^{5/4}
\left(\frac{100GeV}{m_{s}}\right)^{1/8}
\left(\frac{10^{9}GeV}{T_{R}}\right)^{1/4}
MeV  \eq}.$$
The lower bound on $T_{R}$ from requiring the
 elimination of the domain walls then becomes
$${T_{R} \gae 5.4 \;
\frac{\lambda_{s}^{1/2}}{k^{2} \lambda_{a}^{4}}
\left(\frac{m_{s}}{100GeV}\right)^{7/2}
\left(\frac{0.1GeV}{T_{rel}}\right)^{3}
\left(\frac{M_{X}}{10^{10}GeV}\right)^{2}
GeV   \eq}.$$ 
For example, this allows a range of reheating temperatures from
 $T_{R} \approx 5 \times 10^{6}GeV$ up to the gravitino bound to be 
compatible with elimination of the domain walls at $T_{rel} \approx 0.1GeV$
when $\lambda_{a}$ is in the range 0.01 to 1, $k \approx 0.1$
and $M_{X} \approx 10^{10}GeV$.

                 Thus we see that, with $\lambda_{a} \approx k \approx 0.1$, 
$M_{X} \approx 10^{10}GeV$ and 
$T_{R} \approx 10^{7-9}GeV$, the $S$ condensate can
evade thermalization, 
can safely decay before nucleosynthesis (at a 
temperature of around 10MeV) and can 
eliminate the NMSSM domain walls, without 
requiring any small renormalizible couplings. 

         We should note that 
the preferred order of magnitude for $M_{X}$ in this model is 
around $10^{10}GeV$. Values smaller 
than this make it difficult to satisfy the 
no thermalization constraint, equation (51), 
whilst values larger than this make it 
difficult to eliminate the domain walls, 
equation (61), and to ensure that the condensate 
decays before nucleosynthesis, equation (43). 

       So far we have not explained why it was 
necessary to introduce the second discrete symmetry $Z_{B}$. 
In the absence of this discrete symmetry, 
we could have had an additional 
term in the non-renormalizible superpotential,
$${   \frac{\lambda_{b}}{M_{Pl}}S^{3}X    \eq},$$
where $\lambda_{b} \approx 1$.
On integrating out the $X$ field, 
this would give an effective operator in the superpotential, 
$${   \frac{\lambda_{a} \lambda_{b}}{M_{X} M_{Pl}}S^{4}N    \eq},$$
 which lifts the flat direction much earlier 
than the $S^{6}$ superpotential term, such that
$${ S_{o} \approx
 \left(\frac{m_{s}M_{X}M_{Pl}}{\lambda_{a} \lambda_{b}}\right)^{1/3}  \eq}.$$
As a result, the lower bound on $T_{R}$ 
from the requirement that the domain walls
can be eliminated during radiation domination would become
$${ T_{R} \gae 3 \times 10^{12}  \frac{\lambda_{b}^{2/3}}{k \lambda_{a}^{4/3}}
\left(\frac{m_{s}}{100GeV}\right)^{4/3}
\left(\frac{0.1 GeV}{T_{rel}}\right) 
\left(\frac{M_{X}}{10^{10}GeV}\right)^{1/3} GeV 
 \eq}.$$ 
Even with $k \approx \lambda_{a} \approx 1$ 
this would require that $M_{X} \lae 10^{2}GeV$ 
in order to allow reheating temperatures below the gravitino bound, 
making it impossible to avoid thermalizing the condensate. 

                  Thus,
 with non-renormalizible couplings of $S$ to the NMSSM fields,
the $Z_{3}$ domain wall
problem of the NMSSM can be solved even if the renormalizible 
couplings are large and
the reheating temperature is not 
small compared with the gravitino upper bound.
The introduction of 
an intermediate mass scale of the 
order of $10^{10}GeV$ is essential 
for the solution of the domain wall problem in this case, 
since it provides non-renormalizible operators 
which are strong enough to enable
the oscillating scalar field to decay fast enough to avoid
problems with nucleosynthesis and to allow it 
to provide a sufficiently large pressure 
to eliminate the NMSSM domain walls whilst still being weak enough
to prevent thermalization of the discrete 
symmetry breaking scalar condensate before 
the NMSSM domain walls have formed. 
\newpage {\bf \sec. Conclusions.}

              We have considered the possibility of 
solving the NMSSM $Z_{3}$ domain
wall problem by spontaneous discrete symmetry 
breaking occuring during inflation. For the case of
renormalizible couplings of the discrete symmetry 
breaking scalar $S$ to the NMSSM sector
 we find that, if the $S$ scalar has a negative hidden 
sector mass squared term, then it is possible to solve the 
NMSSM domain wall problem without requiring extremely
small couplings so long as the reheating 
temperature is not very large compared 
with that of the electroweak phase transition. 
 Reheating temperatures larger than $10^{7}GeV$ 
would require the couplings to be 
less than $10^{-5}$.
A solution with a positive hidden sector mass squared for $S$ always requires 
very small couplings and so seems generally unnatural.

      For the case of non-renormalizible, Planck-scale suppressed operators, 
a negative hidden sector mass squared solution appears generally ruled out by
 the generation of a large SUSY mass for the $N$ field of the NMSSM. 
For the case of a positive hidden sector mass
 squared, we find that it is possible to solve 
the NMSSM domain wall problem via a 
coherently oscillating $S$ scalar without any small couplings, even 
if the reheating temperature after inflation is large. 
This requires the introduction of an intermediate
mass scale of the order of $10^{10}GeV$, 
in order to allow the $S$ scalars to 
decay without cosmological problems,
and an additional discrete symmetry in order 
to control the allowed non-renormalizible 
terms and so avoid suppressing the density of 
coherent $S$ scalars. 

                Although we have concentrated on the
$Z_{3}$-symmetric NMSSM, the coherently oscillating scalar
 mechanism for eliminating weak
scale domain walls in SUSY models should have rather general 
applications to any SUSY model with a weak scale domain wall
problem, for example, SUSY models with spontaneous CP
violation or spontaneous R-parity violation at the 100GeV to 1TeV scale. 
In these cases, given the sensitivity of the strong CP parameter 
$\overline{\theta}$ to explicit 
CP violating terms and the baryon number violation rate 
to explicit R-parity breaking terms, it may
well be advantageous to have a discrete symmetry breaking expectation value 
which vanishes at zero
temperature, as happens in the case of the coherently 
oscillating scalar mechanism. 

                   This research 
was supported by a European Union Marie Curie 
Fellowship under the TMR programme, contract number ERBFMBICT 950567.


\newpage
\begin{thebibliography}{50}

\bibitem{nilles} H.P.Nilles, Phys.Rep. 110 (1984) 1.

\bibitem{gm} G.F.Giudice and A.Masiero, Phys.Lett. B206 (1988) 480.

\bibitem{nmssm} H.P.Nilles, 
M.Srednicki and D.Wyler, Phys.Lett. 120B (1983) 346.

\bibitem{ets} U.Ellwanger, M. Rausch de Traubenberg and C.A.Savoy,
 Nucl.Phys. B492 (1997) 21.

\bibitem{dwp} Ya.B.Zel'dovich, 
I.Yu.Kobzarev and L.B.Okun, Sov.Phys. JETP 40 (1975) 1,
\newline T.W.B.Kibble, J.Phys. A9 (1976) 1387.

\bibitem{dwp2} A.Vilenkin, Phys. Rep. 121 (1985) 263.

\bibitem{saw} S.A.Abel, S.Sarkar and P.L.White, Nucl.Phys. B454 (1995) 663. 

\bibitem{bag} J.Bagger and E.Poppitz, Phys.Rev.Lett 71 (1993) 2380,
\newline J.Bagger, E.Poppitz and L.Randall, Nucl.Phys. B455 (1995) 59.

\bibitem{abel} S.A.Abel, Nucl.Phys. B480 (1996) 55.

 \bibitem{hm2} M.Dine, L.Randall and 
S.Thomas, Nucl.Phys. B458 (1996) 291.

\bibitem{gravb} J.Ellis, A.Linde and D.Nanopoulos, Phys.Lett. 118B (1982)
59,  \newline M.Yu.Khlopov and A.Linde, Phys.Lett. 138B
(1984) 265,  \newline J.Ellis, J.E.Kim and D.V.Nanopoulos,
Phys.Lett. 145B (1984) 181.

\bibitem{eu} E.W.Kolb
and M.S.Turner, The Early Universe, (Addison-Wesley, Reading 
MA (1990)). 

\bibitem{gravb2} S.Sarkar, Rep.Prog.Phys. 59 (1996) 1493.

\bibitem{cobe} R.K.Schaefer and Q.Shafi, 
Phys.Rev. D47 (1993) 1333; A.R.Liddle, ibid. D49
(1994) 739.

\bibitem{hm} E.Copeland, A.Liddle, D.Lyth, E.Stewart and D.Wands,
Phys.Rev. D49 (1994) 6410,
\newline E.D.Stewart, Phys.Rev.D51 (1995) 6847,
\newline M.Dine, L.Randall and
S.Thomas, Phys.Rev.Lett. 75 (1995) 398.

\bibitem{dti} P.Binetruy and G.Dvali, Phys.Lett. 388B (1996) 241,
\newline E.Halyo, Phys.Lett. 387B (1996) 43.

\bibitem{dj} L.Dolan and R.Jackiw, Phys.Rev. D9 (1974) 3320.

\bibitem{enqrot} K.Enqvist, A.Riotto and I.Vilja, OUTP-97-49-P, hep-ph/9710373.

\bibitem{ad} I.A.Affleck and M.Dine, Nucl.Phys. B249 (1985) 361.

\bibitem{dwp3} J.McDonald, Phys.Lett. 357B (1995) 19. 

\bibitem{dwp4} M.Dine, R.G.Leigh, P.Huet, A.Linde and D.Linde, Phys.Rev. D46 (1992) 
550.

\bibitem{helium} J.Ellis, D.V.Nanopoulos and D.Seckel, Nucl.Phys. B259 (1985) 175.

\bibitem{hadron} M.H.Reno and D.Seckel, Phys.Rev. D37 (1988) 3441. 

\end{thebibliography}
\end{document}